# Digital Forensics Domain and Metamodeling Development Approaches

Omair Ameerbakhsh
College of Computer Science and Engineering. Information system department. Taibah University. Saudi Arabia. Madina
oameerbakhsh@taibahu.edu.sa

Fahad M Ghabban
College of Computer Science and Engineering. Information system department. Taibah University. Saudi Arabia. Madina
fghaban@taibahu.edu.sa

Ibrahim Alfadli
College of Computer Science and Engineering. Information system department. Taibah University. Saudi Arabia. Madina
ialfadli@taibahu.edu.sa

Amer Nizar AbuAli
College of Computer Science and Engineering. Information system department. Taibah University. Saudi Arabia. Madina
aabuali@taibahu.edu.sa

Arafat Al-Dhaqm
School of Computing, Faculty of Engineering University Technology Malaysia
Malaysia, Johor
mrarafat1@utm.my

Mahmoud Ahmad Al-Khasawneh
Faculty of Computer & Information Technology
Al-Madinah International University
Shah Alam, Malaysia
mahmoud@outlook.my

*Abstract*— Metamodeling is used as a general technique for integrating and defining models from different domains. This technique can be used in diverse application domains, especially for purposes of standardization. Also, this process mainly has a focus on the identification of general concepts that exist in various problem domain and their relations and to solve complexity, interoperability, and heterogeneity aspects of different domains. Several diverse metamodeling development approaches have been proposed in the literature to develop metamodels. Each metamodeling development process has some advantages and disadvantages too. Therefore, the objective of this paper is to provide a comprehensive review of existing metamodeling development approaches and conduct a comparative study among them-eventually selecting the best approach for metamodel development in the perspective of digital forensics.

*Keywords*— Digital forensics, Metamodel, Metamodeling

I. INTRODUCTION

Digital Forensics (DFs) domain is a diverse and ambiguous domain. It has many overloading concepts, terminologies, processes, tasks, activities, subdomains, etc [1]–[6]. This diversity and ambiguity make it complex and unstructured among domain forensic investigators[7]–[9]. For example, the database forensics field has many subdomains (e.g: oracle database forensics, MSSQL server database forensics, MySQL database forensics, DB2 database forensic, PostgreSQL database forensics, and SQLite database forensics) which produced many and overloading forensics models and framework for database forensics field [10], [11]. Similarly, mobile forensics, network forensics, IoT forensics, Drone's forensics, email forensics, and wireless forensics [1], [12]–[14]. This diversity makes the DFs field unorganized, ununified, unstructured, complex, heterogeneous, and ambiguous [9], [15]. Therefore, semantic metamodeling language (high abstract model) is required to organize, unify, and structure the DFs domain knowledge in one standardized model. For this purpose, the metamodeling approach is a proper method to develop a sematic metamodeling language (metamodel) for the DFs domain. Metamodeling is used as a general technique for integrating and defining models from different domains [16]. Common concepts of these different views can be identified and shared. The metamodeling technique consequently can be applied in quite different application domains, especially for standardization purposes. Metamodeling is simply the identification of general concepts that exist in each problem domain and their relations. It is used to solve the complexity, interoperability, and heterogeneity of the domain [17], [18]. Metamodels should therefore be rigorously defined as well as being well-structured. The metamodel is a model about a model; it is the explanation of the model. It can specify concepts, attributes, operations, and associations to model a specific domain [19][20]. A metamodel is a precise definition of modeling elements (concepts, attributes, operations, and associations, and rules) needed for creating semantic models [21]. These elements are used to construct a domain model. Also, a metamodel is thus a prescriptive/description model of a modeling language. It is used to solve the ambiguity and heterogeneity of complex domains through the generation of solution models [9], [22], [23].

Therefore, the objective of this paper is to make a comparative analysis among metamodeling approaches to select the best one which may use for organizing and structuring the DFs domain. The results show that the

This paper is organized as follows: Section 1 introduced a brief introduction about metamodeling approaches and DFs domain, the metamodeling development approaches are presented in Section 2. Section 3 provided a comparative analysis among metamodeling approaches, whereas Section 4 displayed the demonstration of the metamodeling approaches, and finally Section 5 summarized this paper.

## II. METAMODELING DEVELOPMENT APPROACHES

Metamodeling is used as a general technique for integrating and defining models from different domains [16]. Common concepts of these different views can be identified and shared. Mainly, this technique can be applied in quite different application domains, especially for standardization purposes. Metamodeling means the identification of general concepts that exist in each problem domain and their relations. It is used to solve the complexity, interoperability, and heterogeneity of the domain [17][18]. Metamodels should therefore be rigorously defined as well as being well-structured. The metamodel development process is used to construct a metamodel, where the process of constructing a metamodel at the M2 level is termed 'metamodel development' [24]. Consequently, the metamodeling development process ensures that the outcome of a metamodel is complete and consistent [25]. Each metamodel development process has advantages and disadvantages. For example, Polynomial regression method PR [26], Finite state machine [27][28], AIMS [29], and Learning-by-doing Approach-Knowledge-based engineering [30] are more suitable for the simulation models. While Kriging [31], Specification-Driven Development of an Executable Metamodel in Eiffel [32]), Towards Automated Testing of Abstract Syntax Specifications of Domain-Specific Modelling Languages [33], Test-driven Approach-model development [34], Metamodeling for Business Model Design [35], and Metamodeling Creation process [24]. Table I displays the comparison among existing metamodeling development approaches.

TABLE I. EXISTING METAMODELING APPROACHES

| Development Process | Description |
|---|---|
| Adaptive and Interactive Modelling System (AIMS) [29] | AIMS is viewed as a learning activity and inductive machine learning techniques from Artificial Intelligence and combined with traditional optimization methods to form a model building system. This metamodeling process includes two steps:<br>1. Competitive Relation Learner (CRL): responsible for generating metamodels from training examples.<br>2. Induction/Selection Optimizer (ISO): uses a multiple-objective optimization method to choose the relevant modeling strategies |
| Polynomial Regression (PR) [26] | PR has been applied in designing complex engineering systems. Originally this polynomial modeling method was developed to produce smooth approximation models of response data contaminated with random error found in the typical physical (stochastic) experiment. It includes two steps:<br>1. Recognize the centrality of diverse outlines considers straightforwardly from the coefficients in the standardized relapse model. For issues with an extensive measurement, it is essential to utilize straight or second-request polynomial models to limit the outline variables to the most basic ones.<br>2. Optimization, the smooth ability of polynomial regression permits a speedy meeting of boisterous capacities. |
| Blind Kriging–engineering design [31] | This process aims to define a metamodel for corporate real estate management it has so far been mainly based on more static approaches such as balanced scorecard, resulting in rather static management models and principles that are needed but inadequate to reflect the dynamic, agile, networked environment of today. It consists of four steps:<br>1. Background review on the topic also on related fields.<br>2. Search for and assemble related studies.<br>3. Gather and code data from studies.<br>4. Builds a framework with the data separated. |
| Specification-Driven Development of an Executable Metamodel in Eiffel [32] | The authors combined specifications and tests to guide the construction of Eiffel metamodels. Specifications are given as Eiffel contracts, whereas tests are written using the acceptance test framework for Eiffel. It consists of five steps:<br>1. A brief modeling phase, where determined the classes that were needed for representing the metamodel.<br>2. Sketch of parts of these class diagrams<br>3. Determine a preliminary set of Eiffel classes.<br>4. Capture a set of well-formed rules in the class diagrams.<br>5. Apply validation and transformation. |
| Learning-by-doing approach-Knowledge-based engineering [30] | Garcia has developed a technology approach that relies on the integration of an object-oriented programming environment and a geometric modeler. The technology has been intensively used by large aerospace and automotive companies to automate repetitive and slightly variant engineering design tasks, thus providing significant results in the design time reduction. It consists of four steps:<br>1. Investigation of the information codes.<br>2. Analysis of general code structure to characterize a bound together deterioration Schema.<br>3. Analysis of the individual learning code items is utilized to characterize the elements in the operation metamodel.<br>4. Investigate the legitimacy of the model. |
| Towards Automated Testing of Abstract Syntax Specifications of Domain-Specific Modeling Languages [33] | This approach is used to support the specification of positive and negative example models from which test models for meta-model testing are generated. The author is especially concerned with the testing of metamodels. |
| Finite State Machine (FSM)-Model-based development [27] | A semantic framework based on Abstract State Machines (ASM) has been offered, which also includes three translational semantics techniques: semantic mapping, semantic hooking, and semantic meta-hooking. However, the author does not demonstrate any tool generation from their semantics specifications. This process consists of:<br>1. Reveals the modeling components for indicating a model of conduct made from a |

| | | |
|---|---|---|
| | | limited number of states, moves between those states, and occasions.<br>2. Creates an "output event" based on its present state and info. One of the states is picked as an underlying state. The depiction of both "deterministic and non-deterministic" (for every pair of state and info occasion there might be a few conceivable next states) FSMs. |
| Test-driven Approach–model development [34] | The authors assigned test cases to the MetaClasses in a meta-model. Test cases are executable models written in PHP and perform transformation like code generation. If a test case shows that a meta-model is inadequate, this must be manually modified. It consists of six steps:<br>1. Recognize domain concepts and their relationship between concepts.<br>2. Improve the metamodel.<br>3. Compose – a test model.<br>4. Execute – the test model.<br>5. Assess (casual) – if success goes to (2) if not go to (6).<br>6. Recognize refactoring (casual). | |
| Metamodeling for Business Model Design [35] | Metamodeling process offered by Hauksson and Johannesson to develop artifact for Business Model Canvas. It consists of 5 steps:<br>1. Explicate problem.<br>2. Outline artifact and define requirements.<br>3. Design and develop artifact.<br>4. Demonstration<br>5. Evaluation | |
| Metamodeling Creation Process [24] | Othman et al., provided metamodeling process creation to develop and validate a domain model for domain knowledge. It consists of 8 steps:<br>1. Models' collection and preliminary domain study.<br>2. Identifying subsets of models to suit research tasks.<br>3. Extraction of general concepts.<br>4. Short-listing candidate definitions.<br>5. Reconciliation of definitions.<br>6. Designation of concepts.<br>7. Identification of relationships.<br>8. Validating the metamodel. | |

III. COMPARATIVE ANALSYSIS

The comparison among metamodeling approaches shows clearly that [24] approach is more suitable for modelling any complex knowledge domain because it is the most recent and also covered whole existing development process steps in other metamodeling process approaches (e.g.: *identify domain source, extract domain concepts, filtering domain concepts, reviewing domain concepts, merging domain concepts, identify domain concepts relationships, design metamodel, validate metamodel, and enhance metamodel*). Also, it has an additional step that requires the researcher to select the most suitable domain model by using coverage measure, as well as a validation step to ensure the correctness and completeness of the metamodel developed. Figure 1 displays the metamodeling development process proposed by [24].

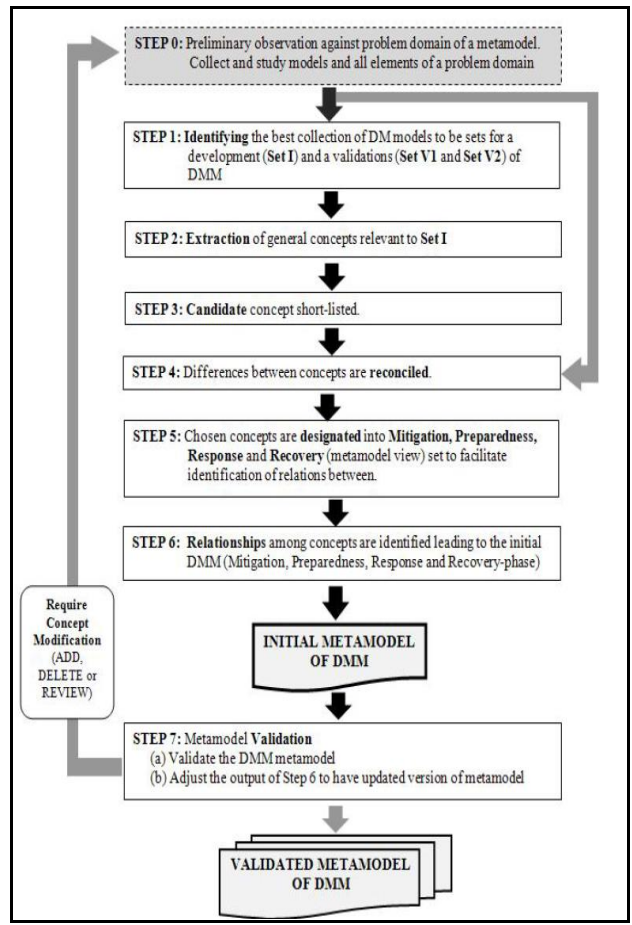

*Fig. 1.* Metamodeling development approach [24]

IV. ORGANIZED AND STRUCTURED DIGITAL FORENSICS DOMAIN USING METAMODELING APPROACH

This section takes a DFs domain as a case study of the complex and heterogeneous domains. As mentioned in Section I, several and overlapping studies exists in literature that is focused in DFs domain. For example, the database forensics field received many works from different authors to deal with different database incidents[36][37] [3], [10], [11], [38]–[46]. However, it lacks the high abstract model. Mobile forensics field received several works to solve mobile incidents [47]–[50]. IoT forensics received several works[51] [52]–[61]. Similarly, several studies [12], [13], [62]–[64] have explored Network forensics-related works. A closely related subdomain of network forensics, cloud forensics, has also received significant research works [1], [8], [53], [65]. Through this of DFs domain, the metamodeling approach is essential for purposes of developing sematic metamodeling language as shown in Figure 2. The M2-Digital forensic metamodel is the highest level of the DFs metamodel which represents the common DFs concepts, M1-Digital forensic models represent the second level/layer of the is DFs metamodel, which governed by M2 level. M0-Digital forensic real models are an instance on the M1-Digital forensic metamodel.

## V. CONCLUSION

This paper reviewed and compared several metamodeling approaches used to structure and organize heterogeneous and complex domains. The best metamodeling approaches have been identified to structure and organize heterogeneity and ambiguity domains. The DFs domain is discussed in this paper as a case study of the heterogenous and complex. domain. Also, this study, suggests developing a semantic metamodeling language for the DFs domain to facilitate, organize, unify, and ruse it among domain users. The future work of this paper is to develop and validate the DFs metamodel using the metamodeling approach.

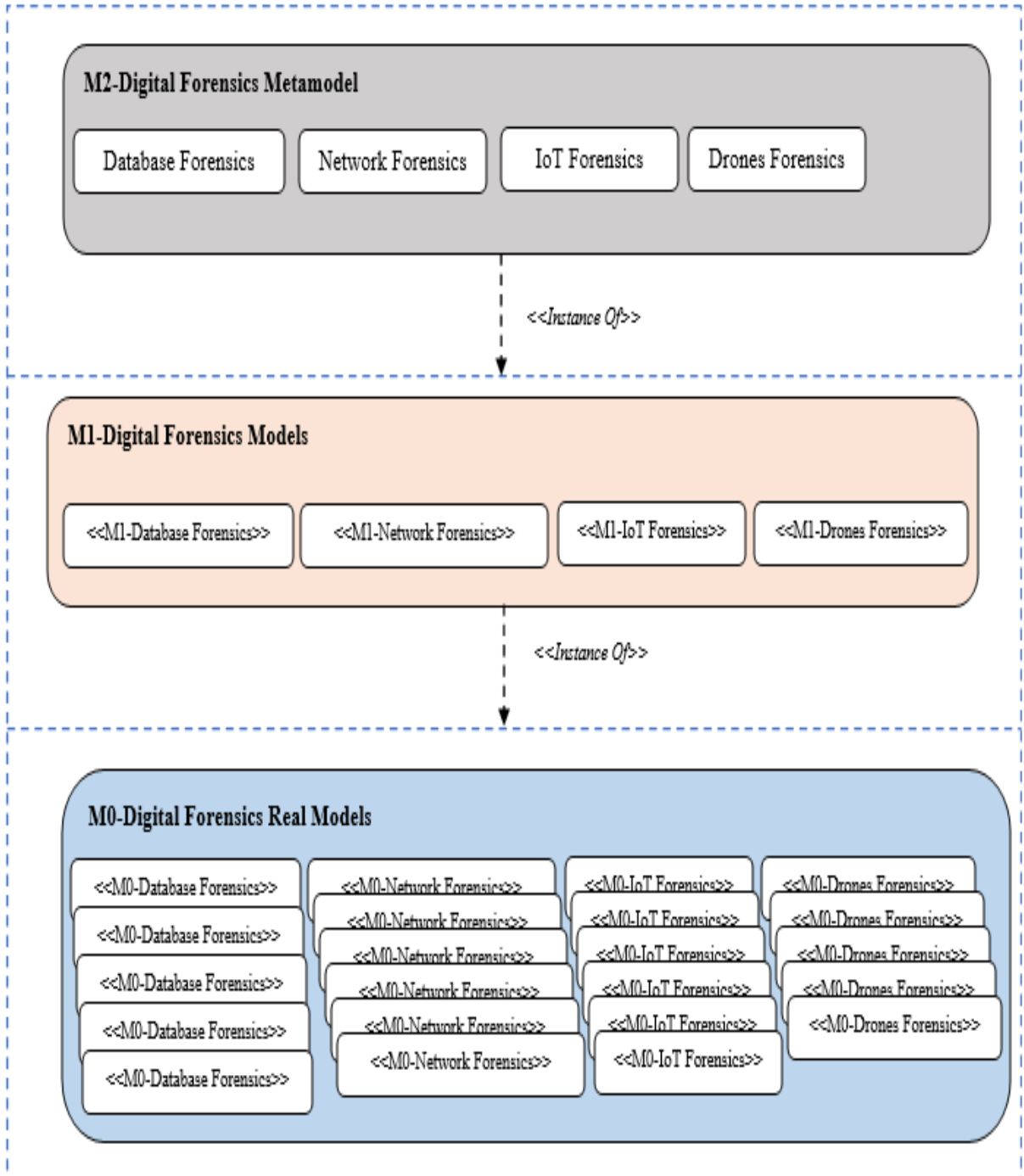

*Fig. 2.* Semantic metamodeling language for DFs domain